\documentclass[12pt]{article}
\usepackage{latexsym}
\usepackage[cp1250]{inputenc}

\title{On threshold amplitudes II: Amplitudes nullification via classical symmetries}
\author{Joanna Domienik\thanks{supported by the {\L}\'od\'z University grant No. 269.},\\ 
 Piotr Kosi\'nski\thanks{supported by KBN grant No. 5 P03B 060 21} \\
Department of Theoretical Physics II \\
University of {\L}\'od\'z \\
Pomorska 149/153, 90 - 236 {\L}\'od\'z/Poland.}
\date{}

\begin{document}
\maketitle
\begin{abstract}
The Ward identities for amplitudes at the tree level are derived from symmetries of the corresponding classical dynamical 
systems. The result are aplied to some $2\rightarrow n$\ amplitudes
\end{abstract}

\newpage
\section{Introducion}

Threshold amplitudes  provide one of the rare examples of problems in quantum field theory which are both interesting 
and tractable, to some extend at least, by analytical methods. They were subject of numerous studies ( see \cite{b1} for 
a review).

An interesting aspect of the problem is amplitude nullification which exhibit some models \cite{b2}. In particular, 
Libanov et al. \cite{b3} have shown that tree amplitudes vanish on the threshold in the model with 0(2)-symmetry 
broken softly by the mass term. Moreover, they gave a beautiful argument in favour of the idea that nullification is 
related to the symmetries of the reduced classical mechanical system \cite{b4}(see also \cite{b5}). The line of reasoning 
presented in Ref. \cite{b4} can be used to construct other theories exhibiting amplitude nullification \cite{b6}. Moreover, 
the conclusions of \cite{b4} were confirmed by more traditional approach based on Ward identities following from the 
symmetry \cite{b7}. In the present note we derive the Ward identities for tree-level threshold amplitudes by functional 
techniques. We assume that the reduced hamiltonian system posses some symmetry and compute the corresponding Ward 
identities at the tree level. These Ward identities are shown to imply amplitude nullification. 

The paper is organized as follows. In Sec. II    we remind  the description of canonical symmetries on lagrangian level. 
Then, in Sec. III, the Ward identities for tree-level amplitudes are computed assuming the existence of symmetry for the 
reduced dynamical system obtained by neglecting space-dependence of the initial system. In Sec.IV these results are 
applied to specific $ 2\rightarrow n$ processes. Finally, Sec.V is devoted to some conclusions.   

\section{Canonical symmetries on lagrangian level}

Generically, the symmetries enforcing amplitudes nullification are canonical rather than point transformations. 
However, it is more convenient to do the pertubation theory on the lagrangian level. Therefore, we shall first remind 
how the canonical symmetries show up in the lagrangian formalism. As  a first step consider a generalized  
lagrangian formalism; the generalization consists in the assumption that the second time derivatives are also
 admitted in the lagrangian function
\begin{eqnarray}
L=L(q,\dot{q},\ddot{q})
\end{eqnarray}
The corresponding equations of motion read
\begin{eqnarray}
\frac{\partial L}{\partial q_i}-\frac{d}{dt}(\frac{\partial L}{\partial \dot  q_i})+\frac{d^2}{dt^2}(\frac{\partial L}
{\partial \ddot q_i})=0
\end{eqnarray}
The class of point transformations allowed is now wider: new coordinates can  depend also on old velocities. The 
transformation $ (q,t)\rightarrow (q',t')$\ is a symmetry, if
\begin{eqnarray}
L(q',\frac{dq'}{dt'},\frac{d^2q'}{dt'^2})\frac{dt'}{dt}
=L(q,\frac{dq}{dt},\frac{d^2q}{dt^2})+\frac{df(q,\frac{dq}{dt},t)}{dt};
\end{eqnarray}
note that $ f$\ can now also depend on $ \frac{dq}{dt}$. Eq.(3) results in the following Noether identity for 
infinitesimal transformations $ q'_i=q_1+\varepsilon y_i, \;\;\; t'=t+\varepsilon \tau , \;\;\; f\rightarrow \varepsilon f$,
\begin{eqnarray}
&&(y_i-\tau \dot q_i)(\frac{\partial L}{\partial q_i}-\frac{d}{dt}(\frac{\partial L}{\partial \dot q_i})+
\frac{d^2}{dt^2}(\frac{\partial L}{\partial \ddot q_i}))+ \nonumber \\
&&+\frac{d}{dt}(L\tau -f+\frac{\partial L}{\partial \dot q_i}(y_i-\tau \dot q_i)+(\dot y_i-\dot \tau \dot q_i-
\tau \ddot q_i)\frac{\partial L}{\partial \ddot q_i}+ \nonumber \\
&&-(y_i-\tau \dot q_i)\frac{d}{dt}(\frac{\partial L}{\partial \ddot q_i}))=0
\end{eqnarray}
Eq. (4) implies the relevant conservation law. The advantage of the generalized formalism is that it gives a more 
general form of Noether symmetries also in the standard case. Assume that L does not depend on $ {\ddot q_i}$'s. 
Then eqs.(3), (4) describe symmetries depending on velocities, even for usual lagrangians. 

Assume now that we have a canonical symmetry described by the generator $ F(q,p)$, i.e. infinitesimally
\begin{eqnarray}
q'_i=q_i+\varepsilon \{q_i,F\}=q_i+\varepsilon \frac{\partial F}{\partial p_i} \\
p'_i=p_i+\varepsilon \{p_i,F\}=p_i-\varepsilon \frac{\partial F}{\partial q_i}
\end{eqnarray}
The symmetry condition reads
\begin{eqnarray}
H(q,p)=H'(q',p')=H(q',p') 
\end{eqnarray}
i.e.
\begin{eqnarray}
\{F,H\}=0
\end{eqnarray}
In order to describe this symmetry on the lagrangian level we express $ p_i$\ in terms of $ \dot q_i$\ through
\begin{eqnarray}
\dot q_i=\frac{\partial H(q,p)}{\partial p_i}
\end{eqnarray}
and define the lagrangian
\begin{eqnarray}
L(q,\dot q)=p_i(q,\dot q)\dot q_i-H(q,p(q,\dot q))
\end{eqnarray}
Consider now the transformation implied by (5)
\begin{eqnarray}
q'_i=q_i+\varepsilon \frac{\partial F}{\partial p_i}(q,p(q,\dot q))
\end{eqnarray}
This is a generalized point transformation in the sense described at the beginning of the present section. Short 
computation using eqs. $ (7)\div (10)$\  gives
\begin{eqnarray}
L(q',\dot q')=L(q,\dot q)+\frac{d}{dt}((p_i\frac{\partial F}{\partial p_i}-F)\mid _{p=p(q,\dot q)})
\end{eqnarray}
which, due to  eq.(4), implies F to be a constant of motion. 

We conclude that the canonical symmetry can be viewed as the generalized point symmetry once the generalized momenta 
are expressed in terms of velocities. 

Note, in particular, that the approach sketched above allows to derive the energy conservation without addressing 
to point transformations with time variation.\\
\section{Ward identities and amplitudes nullification.}
Our starting point is the following lagrangian
\begin{eqnarray}
L=\frac{1}{2}\sum\limits_{i=1}^{k}(\partial \mu \Phi  _i\partial ^{\mu }\Phi  _i-m^2_i\Phi  ^2_i)-V
(\underline{\Phi },\underline{\lambda })
\end{eqnarray}
where $ \underline{\lambda }$\ is the set of coupling constants and $ V(\underline{\Phi }, \underline{\lambda })$\ starts 
from the terms of third order in $ \Phi '_is$. The classical field equations read
\begin{eqnarray}
(\Box +m^2_i)\Phi _i+\frac{\partial V}{\partial \Phi _i}=0
\end{eqnarray}
In order to compute the tree-level amplitudes one can proceed as follows \cite {b8}, \cite{b3}, \cite{b4}, \cite{b5}. 
We look for the solution of the integral equation
\begin{eqnarray}
\Phi _i(x)=\Phi _{0i}(x)-\int d^4y\Delta _{Fij}(x-y)\frac{\partial V}{\partial \Phi _j(y)}
\end{eqnarray}
where $ \Delta _{Fij}(x)=\Delta _F(x;m^2_i)\delta _{ij}$\ is the Feynman propagator, while
\begin{eqnarray}
\Phi _{0i}(x)=\sum\limits_{n}(\beta _i^{(n)}f_{ni}(x)+\overline{\beta }_i^{(n)}\overline{f_{ni}(x)})
\end{eqnarray}
is the solution to the free-field equations ( here $ \{f_{ni}(x)\}_{n=0}^{\infty}$\ is a complete set of normalined 
positive-energy solutions to the corresponding Klein-Gordon equation). Eq.(15) implies eq.(14) together with the 
boundary condition
\begin{eqnarray}
\Phi _i(x)\mid _{\underline{\lambda }=0}=\Phi _{0i}(x)
\end{eqnarray}
but the converse is not true. \\
Once eq.(15) is solved we obtain $ \Phi _i(x)$\ as a function of $ \beta ^{(n)}_i$\ and $ \overline{\beta }_i^{(n)}$. 
Taking derivatives with respect to $ \beta 's$\ and $ \overline{\beta }'s$\ at $ \beta =\overline{\beta }=0$\ one obtains 
matrix elements of field operator between in - and out-states of on -  shell particles in the tree approximation \cite{b5}. 
In order to get the relevant amplitude it remains to reduce, through LZS, the last particle carried by the field 
$ \Phi _i(x)$.  \\
Everything simplifies considerably in the case of threshold amplitudes. Then things become  translational invariant 
and only time dependence remains. The counterparts of $(13)\div (16) $\ read
\begin{eqnarray}
&& L=\frac{1}{2} \sum\limits_{i=1}^{p}( \dot \Phi _i^2-m_i^2\Phi _i^2)-V(\underline{ \Phi }; \underline{\lambda }) \\
&& \Phi_i(t)=\Phi _{0i}(t)-\int\limits_{- \infty}^{ \infty}dt'\Delta_{Fij}(t-t')\frac{ \partial V}{ \partial \Phi_j(t')}\\ 
&& \Phi_{0i}=\beta_i e^{-im_it}+\overline{\beta_i}e^{im_it}
\end{eqnarray}

Now, consider the threshold amplitudes under the assumption that the initial and final particles are of different kinds. 
Then (20) can be replaced by
\begin{eqnarray}
\Phi _{0i}(t)=\left\{\begin{array}{ccl}
\beta _ie^{-im_it},  &  if &  "i" \;is \;an \;initial\; particle \\
\overline{\beta _i}e^{im_it},  & if & "i"\; is\; a \;final\;   particle
\end{array}\right.
\end{eqnarray}
The solution to (19) may not exist or may be not expandable in power series in $ \underline{\lambda }$. To see this 
note that the iterative solution to (19) produces the tree-graph expansion. However, the latter may be not well-defined 
due to vanishing denominators of some propagators.This can be cured if we assume eq.(21) to hold. Indeed, assume 
 the masses $ m_i$\ are such that no combination $ \sum\limits_{i=1}^pn_im_i, \;\; n_i$\ being integers, vanishes. 
Then no propagator   becomes divergent and the solution to (19), viewed as, at least, formal series in $ \underline{\lambda }$, 
exists and is unique; the expansion coefficients are given by the sums of relevant tree-graphs. This series may be 
divergent due to the problem of "small denominators" but this fact is irrelevant for us. Note also that our assumption 
concerning masses $ m_i$\ implies that the free counterpart of (18) $ (\underline{\lambda }=0)$\ has no global 
integrals of motion except individual energies of oscillators ( is not superintegrable). 

Assume now that the theory defined by (18) exhibits some symmetry. We assume the generic situation that the symmetry 
is the canonical transformation rather than less general point transformation. As explained in Sec.II, this means that at 
the lagrangian level, we allow for transformations depending on velocities. So, the infinitesimal form of the symmetry reads
\begin{eqnarray}
\Phi '_i=\Phi _i+\alpha g_i(\underline{\Phi },\underline{\dot \Phi },\underline{m}, \underline{\lambda })
\end{eqnarray}
where $ \alpha $\ is an infinitesimal parameter.\\
Furthermore, we make the following additional assumptions:\\
(i) eq.(22) is a symmetry for generic values of the masses $ m_i$; \\
(ii) the $ \lambda $\ -independent part of eq.(22) reads
\begin{eqnarray}
g_i(\underline{\Phi },\underline{\dot \Phi };\underline{m},\underline{\lambda }=0)=d_i(\underline{m})\dot \Phi _i
\end{eqnarray}
and contains all linear terms. Actually, (ii) is almost a consequence of (i). Indeed, in the limit of vanishing 
interaction we are dealing with the set of decoupled harmonic oscillators with generic frequencies. They have no globally 
defined integrals of motion except individual energies of oscillators. Therefore, the generator of symmetry 
transformations must be a function of them and linearity of the transformation implies that it is a linear combination 
of energies. Using eq. (21) one can translate (23) into the transformation property of $ \beta 's$:
\begin{eqnarray}
&& \beta '_i=\beta _i-i\alpha m_id_i(\underline{m})\beta _i \\   
&& \overline{\beta }'_i=\overline{\beta }_i+i\alpha m_id_i(\underline{m})\overline{\beta }_i \nonumber
\end{eqnarray}
The solution to (19) and (20) will be denoted below by $ \Phi _i(x\mid \beta ,\overline{\beta }$. The uniqueness of the 
solution (viewed as a formal power series) implies the following identity (again, in terms of formal power series) 
resulting from the symmetry (22) \cite{b4}.
\begin{eqnarray}
\Phi '_i(x\mid \beta ,\overline{\beta })=\Phi _i(x\mid \beta ',\overline{\beta }')
\end{eqnarray}
Using (22) and (24) we can rewrite  this identity to the first order in $ \alpha $; then we get 
\begin{eqnarray}
g_i(\Phi (t),\dot \Phi (t);\underline{m},\underline{\lambda })=i\sum\limits_{k-initial}(-\beta _km_kd_k\frac
{\partial \Phi _i(t)}{\partial \beta _k})+i\sum\limits_{k-final}\overline{\beta }_km_kd_k\frac
{\partial \Phi _i(t)}{\partial \overline{\beta }_k}
\end{eqnarray}
This is the basic identity to be used below.\\
Let us apply to both sides the operator 
\begin{displaymath}
 \prod_{k-initial}\frac{\partial ^{n_k}}{\partial \beta _k^{n_k}}\prod_{l-final}\frac{\partial ^{n_l}}{\partial 
\overline{\beta }_l^{n_l}}, 
\end{displaymath}
put $ \beta =\overline{\beta }=0$\ and then apply the K-G operator $ \partial ^2_i+m_i^2$. The result reads 
\begin{eqnarray}
&&(-i)\prod_{k-initial}\frac{\partial ^{n_k}}{\partial \beta _k^{n_k}}\prod_{l-final}\frac{\partial ^{n_l}}{\partial \overline{
\beta }_l^{n_i}}((\partial ^2_t+m_i^2)g_i(\Phi (t),\dot \Phi (t);\underline{m},\underline{\lambda }))\mid _{\beta =
\overline{\beta }=0}= \\
&&=(\sum\limits_{p-initial}(-n_pm_pd_p)+\sum\limits_{r-final}n_rm_rd_r)\prod_{k-initial}\frac{\partial ^{n_k}}
{\partial \beta _k^{n_k}}\prod_{l-final}\frac{\partial ^{n_l}}{\partial \overline{\beta }_l^{n_l}}((\partial _t^2+
m_i^2)\Phi _i(t))\mid _{\beta =\overline{\beta }=0} \nonumber
\end{eqnarray}
Let us now vary the masses $ m_i$\ in such a way that the relation $ \sum\limits_{i}n_im_i=0$\ is allowed for some 
integers $ n_i$. Consider the set of all integer solutions to 
\begin{eqnarray}
\sum\limits_{i}n_im_i=0
\end{eqnarray}
with not all $ n_i$\ vanishing. Call $ \{\overline{n}_i\}$\ the solution to (28) with the following properties: 

(a) $\sum\limits_{i}\mid n_i\mid $\ attains minimal value 

(b) the amplitudes with $ \mid \overline{n}_i\mid $\ external lines corresponding to i-th kind of particles are a priori 
allowed by the topology of the graphs implied by the lagrangian (18) 

Now we can make precise which kinds of particles are viewed as initial and final ones. Namely, if $ \overline{n}_i>0 \;\;
(\overline{n}_i<0)$, the "i"-th kind of particles will be called initial (final) (one can forget about those particles 
for which $\overline{n}_i=0$\ ).\\
Let us now consider eq.(27) with $ n_k=\mid \overline{n}_k\mid ,  \;\; k\neq i$\ and $ n_i=\mid \overline{n}_i\mid -1$. 
We multiply it by $ exp(\mp im_it)$, depending on whether "i"-th particle is initial or final, and integrate over t. 
It is easy to see that, due to our assumptions, everything here is perfectly finite-the only divergence in propagator, 
emerging due to the relation $ \sum\limits_{k}\overline{n}_km_k=0$, is killed by K-G operator. One can say even more. 
On the right-hand side one gets the amplitude $ A(\overline{n}\rightarrow \overline{n})$\ describing the production of 
final particles out of inital ones, all of multiplicities $ \mid \overline{n}_k\mid $, multiplied by the factor 
$( -\sum\limits
_{p}\overline{n}_pm_pd_p\pm m_id_i)$\ (the sign depends on whether "i"-th particle is initial or final). On the left-hand 
side only the linear term in $ g_i$\ survives. Taking into account energy conservation and the form (23) of linear part 
of $ g_i$\ we conclude that the left-hand side equals $ \pm m_id_iA(\overline{n}\rightarrow \overline{n})$. Collecting 
all that we arrive at the following identity
\begin{eqnarray}
(\sum\limits_{p}\overline{n}_pm_pd_p)A(\overline{n}\rightarrow \overline{n})=0
\end{eqnarray}
which implies $ A(\overline{n}\rightarrow \overline{n})\neq 0$\ only provided
\begin{eqnarray}
\sum\limits_{p}\overline{n}_pm_pd_p=0
\end{eqnarray}

The latter, however, does not hold in general. \\
Let us reconsider the examples described in Ref.\cite {b4}. For the 0(2)-symmetric model broken softly by the mass term 
one has two kinds of particles with $ d_1=0, \;\; d_2=\frac{1}{2}(m_1^2-m_2^2)$. Note that in this case the identity (19) 
has been also proven diagramatically in \cite {b7}. Another model considered in Ref.\cite{b4} is the Henon-Heiles system 
in its integrable version. Again, we are dealing with two kinds of particles with $ d_1=\frac{1}{2}(4m_1^2-m_2^2), \;\;
d_2=0$.

 In general our basic identity allows us to check in each case whether a given symmetry implies amplitude 
nullification.
\section{$ 2\rightarrow n$\ in 0(N)-symmetric models.}
In order to find more interesting application of our Ward identities let us consider the 0(N) model defined by the 
Lagrangian
\begin{eqnarray}
L=\frac{1}{2}\sum\limits_{i=1}^N(\partial _{\mu }\Phi _1\partial ^{\mu }\Phi _1-m^2\Phi _1^2)-\frac{\lambda }{4!}
(\sum\limits_{i=1}^{N}\Phi _i^2)^2
\end{eqnarray}
Consider the production of $ n$\ particles at the threshold by two particles with three-momenta $ \vec p$\ and 
$ -\vec p, \;\; \vec p\neq 0$; we assume that both initial particles are carried by $ \Phi _1$\ while the final 
ones correspond to other indices. The algorithm to calculate such an amplitude at tree level is well known \cite {b9}. 
First, we solve field equations for the fields $ \Phi _i, \;\; i\neq 1$, with the boundary conditions:
\begin{eqnarray}
&& \Phi _1\equiv 0 \nonumber \\
&& \Phi _i\mid _{\lambda =0}=z_ie^{imt},  \;\;\;\;  i\neq 1
\end{eqnarray}
The solution reads (cf. \cite {b8})
\begin{eqnarray}
\Phi _{icl}(t)=\frac{z_ie^{imt}}{1-\frac{\lambda }{48m^2}(\sum\limits_{k\neq 1}z_k^2)e^{2imt}}
\end{eqnarray}
Then the field equation for $ \Phi _1$\ is written out keeping only linear terms in $ \Phi _1$:
\begin{eqnarray}
(\Box +m^2)\Phi _1+\frac{\lambda }{3!}\Phi _1(\sum\limits_{k\neq 1}\Phi _k^2)=0
\end{eqnarray}
Comparying eqs.(33) and (34) with eq.(3.3) of Ref.\cite {b9} we conclude that (34) and (3.3) coincide provided 
the identification a=2, $ z_0^2\rightarrow \sum\limits_{k\neq 1}z^2_k$\ has been made. Then the results of 
Ref \cite {b9} imply that the scattering amplitude for the process $ 2\rightarrow n$\ vanishes  unless n=2 which, 
however, is excluded by energy conservation $ (\vec p\neq 0)$. \\
To find a deeper reason for such nullification consider the tree graph  contributing to the amplitude under consideration. 
The nonvanishing three momentum $ \vec p$\ flows only through the continuous line of propagators connecting both initial 
lines. One can immediately conclude that our amplitude coincides with the one for the process $ 2\rightarrow n$\ with all, 
both initial and final, particles at threshold and the mass squared $ m^2$\ of the first particle replaced by 
$ M^2=m^2+\vec p^2$; the "effective" Lagrangian reads
\begin{eqnarray}
L= \frac{1}{2}\sum\limits_{i=2}^{N}(\partial _{\mu }\Phi _i\partial ^{\mu }\Phi _i-m^2\Phi _i^2)
+\frac{1}{2}(\partial _{\mu }
\Phi _1\partial ^{\mu }\Phi _1-M^2\Phi _1^2)-\frac{\lambda }{4!}(\Phi _1^2+\sum\limits_{i=2}^{N}\Phi _i^2)^2
\end{eqnarray}
When reduced to one dimension, $ \partial _{\mu }\Phi _i\rightarrow \dot \Phi _i$, this becomes the so-called Garnier 
system which is integrable \cite {b10}, the additional, apart from energy, integrals of motion being the generators $ l_{ij}, 
i,j\neq 1$\ of O(N-1) symmetry  together with 
\begin{eqnarray}
F=\frac{\lambda \cdot \sum\limits_{k=2}^{N}l_{1k}^2}{4!(M^2-m^2)}+p_1^2+\frac{\lambda }{4!}\Phi _1^2(\sum\limits_{k=1}^N
\Phi _k^2)
\end{eqnarray}
Note that the transformation generated by F obeys (i), (ii) (as for as (i) is concerned, it can be checked that F, suitably 
modified, continues to be a constant of motion even if the masses of other particles are varying as long as they 
are different from M). In particular,
\begin{eqnarray}
d_i(\underline {m})=2\delta _{i1}
\end{eqnarray}
Therefore, eq.(   ) gives
\begin{eqnarray}
4MA(2\rightarrow n)=0
\end{eqnarray}
For $ \vec p\rightarrow 0$, F (multiplied by $ M^2-m^2$\ ) becomes a function of 0(N) generators and the 
above reasoning breaks down. The process $ 2\rightarrow 2$\ becomes then possible and the relevant amplitude is obviously 
nonvanishing. \\
\section{Conclusions.}
We have derived the basic identity (26) which generates all Ward identities on tree level. Under some further assumptions 
these identities imply vanishing of certain threshold amplitudes. We saw that the important point here is the form of 
coupling constant(s) independent part of symmetry transformations. Indeed, eq.(30) express the conservation law for the 
generator of free-field $ (\underline {\lambda }=0)$\ counterpart of symmetry transformations. Let us recall the general 
field-theoretical framework  of selection rules for scattering amplitudes. Symmetry implies conservation law for some charge 
Q built from interpolating fields and their conjugate momenta. The infinitesimal action  of symmetry on interpolating 
fields is given by their commutator with Q. If this action is linear in fields we can use LZS asymptotic condition together 
with constancy of Q to find the transformation properties of asymptotic fields which, in turn, allows to reconstruct 
$ Q_{in}$\ and $ Q_{out}$; obviously $ Q_{in}=Q=Q_{out}$\ which  implies appropriate selection rule. \\
For nonlinear transformations situation is more involved; the (weak) LZS asymptotic condition is insufficient to derive the 
the asymptotic form of symmetry in such a straightforward manner. In our case it is reflected by the fact that the 
conservation law for free-part of generator holds only in the lowest-energy sector allowed by energy conservation.     
For higher  sectors  nonlinear parts of symmetry transformations survive asymptotics, i,e. the amputation procedure 
(more precisely, the relevant contributions are ambiguous due to divergences in some propagators).

Eq.(30) gives an 
additional ( apart from space-time symmetries) constraint which implies  amplitude nullification. The reasoning made 
here strongly resembles the arguments in favour of the well-known Coleman-Mandula \cite {b11} theorem. There, mixing of 
internal and space-time symmetries implies the existence of conserved charges equipped with space-time indices. Therefore 
the relevant conservation laws give rise to the selection rules involving fourmomenta which contradicts analyticity 
\cite {b11}.\\
In our case the role of space-time and internal symmetries mixing is played by the fact that the symmetry transformations 
involve (even in their linear part)time derivatives of fields which results in the appearance of masses in eq. (30). 
The analogy with Coleman-Mandula theorem is also clearly seen in the case of $ 2\rightarrow n$\ process. As long as 
$ \vec p\neq 0 \;\; (M^2\neq m^2)$\ we have at our disposal an integral of motion with "space-time indices", due to 
appearance of time derivatives in F (throught $ p_1$\ ). In the limit $ \vec p\rightarrow 0$\ the symmetry is 
purely internal and the process $ 2\rightarrow 2$\ becomes possible. \\
The last example is also interesting because it provides an explanation for nullification phenomena in 
some $ 2\rightarrow n$\ processes. It seems that the trick consisting in replacing our process with the one with all 
particles at rest at the price of modifying the masses of initial particles is more general at least for theories 
with scalar particles only. One has to embedd the relevant dynamics into the one described by lagrangian with 
a symmetry which forcces some of the amplitudes to vanish.

\end{document}